\documentclass[11pt]{article}
\usepackage[margin=1in]{geometry}
\usepackage{amsmath}
\usepackage{amssymb}
\usepackage{algorithm}
\usepackage{algpseudocode}
\usepackage{graphicx}
\usepackage{booktabs}
\usepackage{microtype}
\usepackage{xurl}
\usepackage[font=small,labelfont=bf]{caption}
\usepackage{xcolor}
\usepackage[colorlinks=true,linkcolor=blue!50!black,citecolor=blue!50!black,urlcolor=blue!50!black]{hyperref}

\title{Parallelizing the Variational Quantum Eigensolver: \\
From JIT Compilation to Multi-GPU Scaling}

\author{
R. Malarchick\textsuperscript{1} and A. Steed\textsuperscript{1} \\[0.5em]
\textsuperscript{1}Department of Physical Sciences, Embry-Riddle Aeronautical University, \\
Daytona Beach, FL 32114, USA \\[0.5em]
\texttt{rylan1012@gmail.com, steeda1@my.erau.edu}
}

\date{January 2026; revised October 2026}

\begin{document}

\maketitle

\begin{abstract}
    The Variational Quantum Eigensolver (VQE) is a hybrid quantum-classical algorithm for ground-state energies of molecular systems. We compute the potential energy surface of the hydrogen molecule (H$_2$) at 100 bond lengths with PennyLane on the ERAU Vega cluster (AMD EPYC 9654 CPUs, 4 NVIDIA H100 GPUs) and compare a serial baseline, an Optax and Catalyst JIT version, a \texttt{lightning.gpu} version, and an MPI version. Early stopping, a larger learning rate, and warm starts reduce the iteration count from 30,000 to 7,033 and the runtime from 593.95~s to 143.80~s, while the time per iteration stays between 0.0198~s and 0.0204~s. At 4 qubits, \texttt{lightning.gpu} lowers the time per iteration to 0.0145~s. MPI runs with 2 to 32 ranks finish in 8.45~s to 5.04~s; the serial and MPI jobs ran under different core allocations, so we do not report their ratio as a parallel speedup. In a 4 to 26 qubit study with different CPU and GPU software stacks, the CPU/GPU time ratio ranges from 3.5 to 81.5. One H100 runs up to 29 qubits.
\end{abstract}

\section{Introduction}

Quantum chemistry computes molecular structure, reaction pathways, and material properties from first principles. A central problem is the ground-state energy and wavefunction of a molecule. The cost of an exact calculation grows exponentially with the number of orbitals, which limits exact classical methods to small molecules~\cite{mcardle2020}.

The Variational Quantum Eigensolver (VQE) is a hybrid quantum-classical algorithm~\cite{peruzzo2014,cerezo2021} designed for noisy near-term devices. A parameterized quantum circuit (the ansatz) prepares trial wavefunctions, and a classical optimizer adjusts the parameters to minimize the energy. Peruzzo et al.~\cite{peruzzo2014} first demonstrated VQE on a photonic quantum processor. Since then, ansatz design has been a major focus: the unitary coupled cluster (UCC) ansatz~\cite{romero2018} is now standard for molecular simulations, while hardware-efficient ansatzes~\cite{kandala2017} reduce circuit depth on noisy devices.

Much of the algorithm development for VQE uses classical state-vector simulation, and several software tools speed up this simulation. PennyLane~\cite{bergholm2018} provides automatic differentiation for quantum circuits, and its Lightning backend~\cite{pennylane_lightning} provides compiled state-vector simulation. NVIDIA's cuQuantum~\cite{cuquantum2023} targets multi-GPU simulation, and JAX~\cite{jax2018} enables JIT compilation for numerical code. MPI parallelization of quantum chemistry is well established~\cite{valiev2010}.

We study the hydrogen molecule (H$_2$), the simplest neutral molecule and a standard benchmark for quantum chemistry methods. Its potential energy surface has an equilibrium bond length and a dissociation limit. We ask two questions. First, can VQE with a single-parameter ansatz compute the H$_2$ potential energy surface accurately? Second, how do JIT compilation, GPU simulation, and MPI distribution change the runtime of this scan on one HPC node with four H100 GPUs? We answer the second question with measured runtimes, and we report each runtime together with the optimizer and stopping settings that produced it.

Section~\ref{sec:model} states the problem and the model. Section~\ref{sec:methods} describes the implementations and the planned parallel speedups. Section~\ref{sec:results} reports the measurements, and Section~\ref{sec:discussion} discusses what they do and do not show.

\section{Problem and Model}
\label{sec:model}

We compute the ground-state energy $E_0$ of H$_2$ as a function of the internuclear distance $d$. In the Born-Oppenheimer approximation and in atomic units, the molecular Hamiltonian is
\begin{equation}
    H = -\frac{1}{2}\sum_{i=1}^{2}\nabla_i^2 - \sum_{i=1}^{2}\left(\frac{1}{|\mathbf{r}_i - \mathbf{R}_A|} + \frac{1}{|\mathbf{r}_i - \mathbf{R}_B|}\right) + \frac{1}{|\mathbf{r}_1 - \mathbf{r}_2|} + \frac{1}{d},
\end{equation}
where $\mathbf{r}_i$ are the electron positions and $\mathbf{R}_A$ and $\mathbf{R}_B$ are the nuclear positions. We project this Hamiltonian onto the minimal STO-3G basis~\cite{szabo1996}. In second quantization it reads
\begin{equation}
    H = \sum_{i,j} h_{ij} a_i^\dagger a_j + \frac{1}{2}\sum_{i,j,k,\ell} h_{ijk\ell}\, a_i^\dagger a_j^\dagger a_k a_\ell,
\end{equation}
where $h_{ij}$ are the one-electron integrals (kinetic energy and nuclear attraction), $h_{ijk\ell}$ are the two-electron integrals (electron-electron repulsion), and $a_i^\dagger$ and $a_i$ are fermionic creation and annihilation operators. PennyLane computes these integrals over the Hartree-Fock molecular orbitals of the STO-3G basis and maps the Hamiltonian to Pauli operators on 4 qubits with the Jordan-Wigner transformation.

VQE relies on the variational principle: for any normalized trial wavefunction $|\psi(\theta)\rangle$ with parameters $\theta$, the energy expectation value is bounded below by the ground-state energy,
\begin{equation}
    E(\theta) = \langle \psi(\theta) | H | \psi(\theta) \rangle \geq E_0 .
\end{equation}
The minimum of $E(\theta)$ is an upper bound on $E_0$, and it is close to $E_0$ when the ansatz can represent a state close to the ground state. We prepare the trial state as $|\psi(\theta)\rangle = U(\theta)\,|\text{HF}\rangle$, where $|\text{HF}\rangle = |1100\rangle$ is the Hartree-Fock reference state with both electrons in the lowest spatial orbital, and $U(\theta)$ is the double excitation gate
\begin{equation}
    U(\theta) = \exp\left(-i\frac{\theta}{2}\left(a_0^\dagger a_1^\dagger a_2 a_3 - a_3^\dagger a_2^\dagger a_1 a_0\right)\right).
\end{equation}
This ansatz describes the dominant correlation in H$_2$, the double excitation of both electrons from the bonding to the antibonding orbital, with a single parameter $\theta$. In simulation, the gate is a parameterized unitary applied to the $2^N$-dimensional state vector, and VQE solves the one-parameter problem $\theta^* = \arg\min_\theta E(\theta)$ with a gradient-based optimizer.

\section{Methods}
\label{sec:methods}

\subsection{Workload}

The potential energy surface requires $E(\theta^*)$ at $N_b = 100$ bond lengths, uniformly spaced in $[0.35, 3.0]$~\AA. For each bond length $d_i$, the baseline builds the Hamiltonian $H(d_i)$ with Hartree-Fock, starts from $\theta_0 = 0$ (the Hartree-Fock state), runs 300 Adam steps with learning rate $\alpha = 0.01$, and stores $E_i = E(\theta^*_i)$ (Algorithm~\ref{alg:serial}). Each step evaluates the cost and its gradient. The baseline therefore runs 30,000 optimization steps in total. The bond lengths are independent of each other, so the scan is embarrassingly parallel: $E_i = f(d_i)$ for $i = 1, \ldots, 100$, where $f$ is the VQE optimization.

A state-vector simulator stores $2^n$ amplitudes for $n$ qubits and applies each gate in $O(2^n)$ operations. For H$_2$ the state vector has 16 amplitudes.

\begin{algorithm}[t]
    \caption{Serial VQE baseline for the H$_2$ potential energy surface}
    \label{alg:serial}
    \begin{algorithmic}[1]
        \State \textbf{Input:} bond lengths $\{d_1, \ldots, d_{100}\}$; \textbf{Output:} energies $\{E_1, \ldots, E_{100}\}$
        \State Initialize the \texttt{lightning.qubit} device with 4 qubits
        \For{$i = 1$ to $100$}
        \State Generate $H(d_i)$ with Hartree-Fock (STO-3G basis)
        \State $\theta \gets 0$; initialize Adam with $\alpha = 0.01$
        \For{$j = 1$ to $300$}
        \State $E \gets \langle \psi(\theta) | H(d_i) | \psi(\theta) \rangle$ and $\nabla_\theta E$ (PennyLane default gradient method)
        \State $\theta \gets \text{Adam\_step}(\theta, \nabla_\theta E)$
        \EndFor
        \State $E_i \gets E(\theta)$
        \EndFor
    \end{algorithmic}
\end{algorithm}

\subsection{Implementations}

Table~\ref{tab:implementations} summarizes the four implementations. The JIT version (\texttt{vqe\_serial\_optax.py}) compiles the full optimization loop with Catalyst (\texttt{@qjit}, \texttt{catalyst.grad}, \texttt{catalyst.while\_loop}) and replaces PennyLane's Adam with the Optax Adam optimizer. It also differs from the baseline in three settings: the learning rate is 0.05 instead of 0.01, the loop stops when the energy changes by less than $10^{-8}$~Ha between steps (with at most 300 steps), and each bond length starts from the optimized parameters of the previous bond length instead of from zero. The GPU version (\texttt{vqe\_gpu.py}) uses \texttt{lightning.gpu} with the same optimizer and stopping rule, without Catalyst, and starts every bond length from zero. The MPI version (\texttt{vqe\_mpi.py}) splits the 100 bond lengths into contiguous blocks, one per rank, and each rank runs the JIT runner on its block.

\begin{table}[t]
    \centering
    \small
    \setlength{\tabcolsep}{4pt}
    \begin{tabular}{llllll}
        \toprule
        Version & Device & Compilation & Optimizer (lr) & Stopping rule & Start per bond \\
        \midrule
        Baseline & \texttt{lightning.qubit} & none & PennyLane Adam (0.01) & fixed 300 steps & zeros \\
        JIT & \texttt{lightning.qubit} & Catalyst & Optax Adam (0.05) & $|\Delta E| < 10^{-8}$ & previous \\
        GPU & \texttt{lightning.gpu} & none & Optax Adam (0.05) & $|\Delta E| < 10^{-8}$ & zeros \\
        MPI & \texttt{lightning.qubit} & Catalyst & Optax Adam (0.05) & $|\Delta E| < 10^{-8}$ & previous \\
        \bottomrule
    \end{tabular}
    \caption{The four H$_2$ implementations (scripts \texttt{main.py}, \texttt{vqe\_serial\_optax.py}, \texttt{vqe\_gpu.py}, and \texttt{vqe\_mpi.py}). All versions stop after at most 300 steps per bond length. ``Previous'' means that a bond length starts from the optimized parameters of the preceding bond length (within each rank for MPI).}
    \label{tab:implementations}
\end{table}

All runs used one node of the ERAU Vega cluster with two AMD EPYC 9654 CPUs (192 cores), 1.5~TB of memory, four NVIDIA H100 PCIe GPUs (80~GB each), and an InfiniBand HDR interconnect. The software stack was Python 3.12, PennyLane 0.43.1, PennyLane-Catalyst 0.13.0, JAX 0.6.2, Optax 0.2.6, CUDA 11.8, and OpenMPI 4.x.

\subsection{Planned Speedups}
\label{sec:prediction}

Before the runs, we expected 2--5$\times$ from JIT compilation, from compiling the circuit and the optimization loop, fusing operations, and computing gradients with vectorized operations. For MPI we expected about $0.8p$ for $p$ ranks, below the ideal $p$ because of communication and the JIT compilation on each rank. Amdahl's law~\cite{amdahl1967} gives the strong-scaling speedup with $p$ processors as
\begin{equation}
    S_p = \frac{1}{f_s + f_p/p},
\end{equation}
where $f_s$ and $f_p = 1 - f_s$ are the serial and parallel fractions of the work. With an assumed $f_s = 0.05$ (initialization, I/O, and plotting), the predicted speedups are 3.48, 5.93, 9.14, and 13.56 for $p = 4$, 8, 16, and 40 (Table~\ref{tab:predictions}).

We profiled a serial run with 10 bond lengths and 50 iterations with Python's \texttt{cProfile}. QNode calls (\texttt{qnode.\_\_call\_\_}), which include the gradient computation (\texttt{\_grad\_with\_forward}), take 28.2~s of the 32.7~s runtime (86\%). This measured parallel fraction of 0.86 is lower than the assumed 0.95.

\begin{table}[t]
    \centering
    \small
    \begin{tabular}{rrr}
        \toprule
        Processors & Ideal speedup & Predicted speedup \\
        \midrule
        4  & 4.0  & 3.48  \\
        8  & 8.0  & 5.93  \\
        16 & 16.0 & 9.14  \\
        40 & 40.0 & 13.56 \\
        \bottomrule
    \end{tabular}
    \caption{Speedups predicted by Amdahl's law with an assumed serial fraction $f_s = 0.05$.}
    \label{tab:predictions}
\end{table}

We first considered the Ray framework for its task-based API. The Ray pip package failed to install alongside PennyLane-Catalyst because of conflicting transitive dependencies, and we did not resolve the version constraints. We therefore used MPI through \texttt{mpi4py}. OpenMPI is installed on the Vega cluster, \texttt{mpi4py} installed without conflicts next to PennyLane, Catalyst, and JAX, and the static split of 100 bond lengths needs no dynamic task scheduling.

\section{Results}
\label{sec:results}

\subsection{Potential Energy Surface}

\begin{figure}[t]
    \centering
    \includegraphics[width=0.7\textwidth]{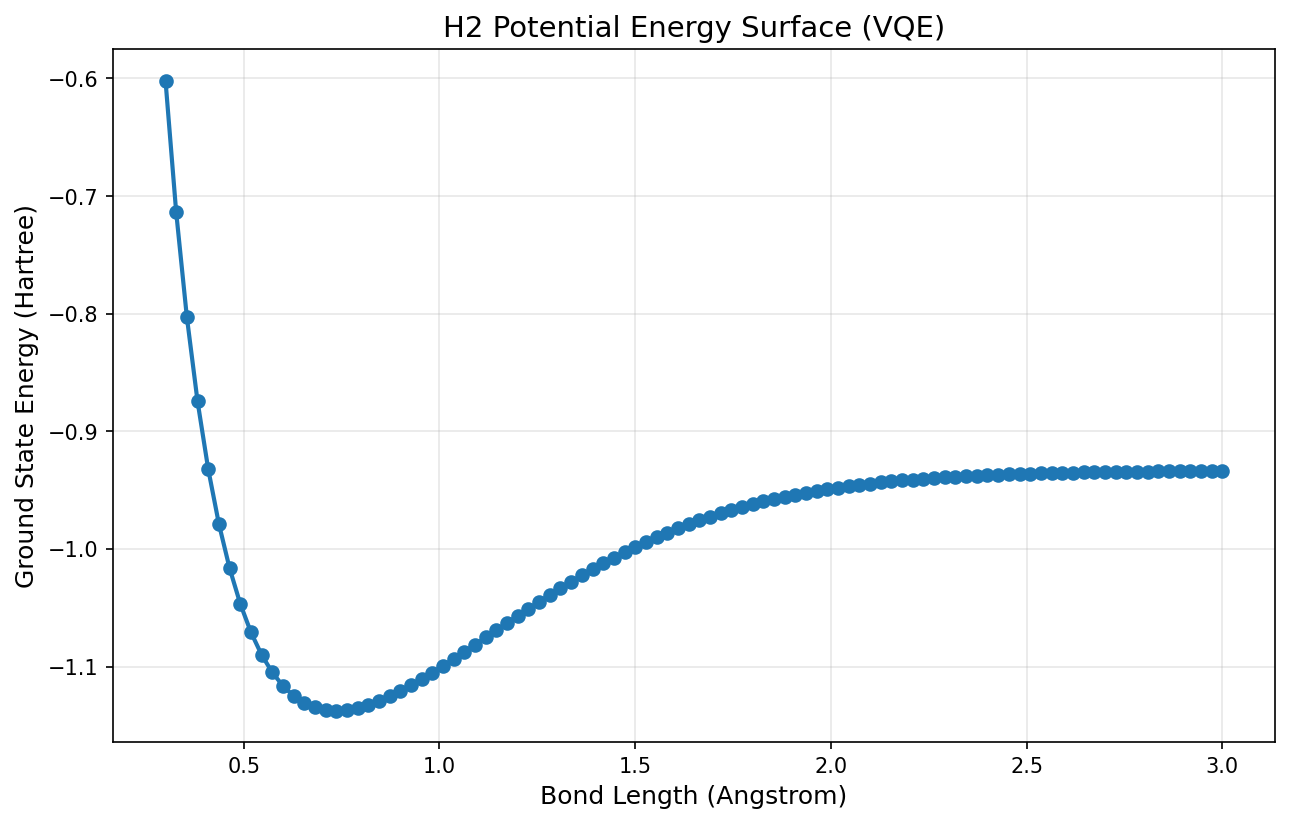}
    \caption{H$_2$ potential energy surface from the serial VQE implementation, with the minimum near 0.74~\AA\ and the approach to the dissociation limit at large bond lengths.}
    \label{fig:serial_pes}
\end{figure}

Figure~\ref{fig:serial_pes} shows the computed potential energy surface. It has a repulsive wall at short bond lengths, a minimum near 0.74~\AA, and an approach to the separated-atom limit beyond 2.5~\AA. The grid values in this subsection come from the JIT run. The lowest energy on the grid lies at 0.725~\AA, and the next grid point is 0.752~\AA\ (grid spacing 0.027~\AA). The experimental bond length of 0.741~\AA\ lies between them. The lowest VQE energy on the grid is $-1.137217$~Ha at 0.725~\AA. Exact diagonalization of the same STO-3G qubit Hamiltonian in the two-electron sector gives $-1.137221$~Ha, a difference of $4\times10^{-6}$~Ha.

In the STO-3G basis, the ground state of H$_2$ lies in the span of the Hartree-Fock state $|1100\rangle$ and its double excitation $|0011\rangle$, because single excitations change the spatial symmetry. The single-parameter ansatz can therefore represent the exact STO-3G ground state at every bond length. Exact diagonalization at 0.35, 0.725, 1.5, and 3.0~\AA\ confirms that the ground state has unit weight in this two-dimensional span.

\subsection{Single-Process Runtimes}

Table~\ref{tab:h2_runtimes} lists the runtimes of the three single-process versions. The JIT version has a 4.13 times lower runtime than the baseline (143.80~s against 593.95~s) and a 4.3 times lower iteration count (7,033 against 30,000). The time per iteration is the runtime divided by the iteration count, and it includes the Hartree-Fock Hamiltonian construction for each bond length. Because this per-bond cost is spread over about 70 iterations per bond length in the JIT run and over 300 in the baseline, the time per iteration does not isolate the cost of the compiled circuit. The measurements show that the runtime reduction tracks the iteration count. We did not change the four settings one at a time, so we cannot divide the reduction among them.

The GPU version has the lowest time per iteration of the three (0.0145~s, against 0.0198~s for the baseline and 0.0204~s for the JIT version). It needs 11,345 iterations, against 7,033 for the JIT version, which starts each bond length from the previous solution, and it therefore finishes later (164.91~s against 143.80~s). The two versions differ in device, compilation, and starting point at the same time, so this comparison does not isolate the effect of the GPU.

\begin{table}[t]
    \centering
    \small
    \begin{tabular}{lrrr}
        \toprule
        Version & Runtime (s) & Iterations & Time per iteration (s) \\
        \midrule
        Baseline & 593.95 & 30,000 & 0.0198 \\
        JIT      & 143.80 & 7,033  & 0.0204 \\
        GPU      & 164.91 & 11,345 & 0.0145 \\
        \bottomrule
    \end{tabular}
    \caption{Single-process runtimes for the 100-bond-length H$_2$ scan. Time per iteration is the runtime divided by the iteration count and includes the Hamiltonian construction for each bond length.}
    \label{tab:h2_runtimes}
\end{table}

\subsection{MPI Runs}

Table~\ref{tab:mpi_results} lists the MPI runtimes for 2 to 32 ranks. Each value is the elapsed time on rank 0, measured before the final gather. Among the MPI runs, the runtime falls from 8.45~s at 2 ranks to 5.04~s at 32 ranks.

We do not report a parallel speedup for these runs. The serial JIT job requested one core (\texttt{ppn=1}), and the MPI jobs ran under separate allocations. The per-bond times of the serial run assign 73.5~s of work to rank 0 and 70.3~s to rank 1 at 2 ranks, while the measured 2-rank runtime is 8.45~s. A balanced split of the same work cannot explain this factor of 8.7, so the serial and MPI jobs had different resources available, for example the number of cores each process could use. The timer also covers rank 0 only. We cannot rerun both jobs under one allocation, so the ratio of their runtimes mixes the effect of the rank count with the effect of the allocation.

\begin{table}[t]
    \centering
    \small
    \begin{tabular}{lrrrrr}
        \toprule
        Ranks & 2 & 4 & 8 & 16 & 32 \\
        \midrule
        Runtime on rank 0 (s) & 8.45 & 6.07 & 5.48 & 5.06 & 5.04 \\
        \bottomrule
    \end{tabular}
    \caption{MPI runtimes for the 100-bond-length scan, measured on rank 0 before the final gather.}
    \label{tab:mpi_results}
\end{table}

\subsection{CPU and GPU Scaling Study}

\begin{figure}[t]
    \centering
    \includegraphics[width=\textwidth]{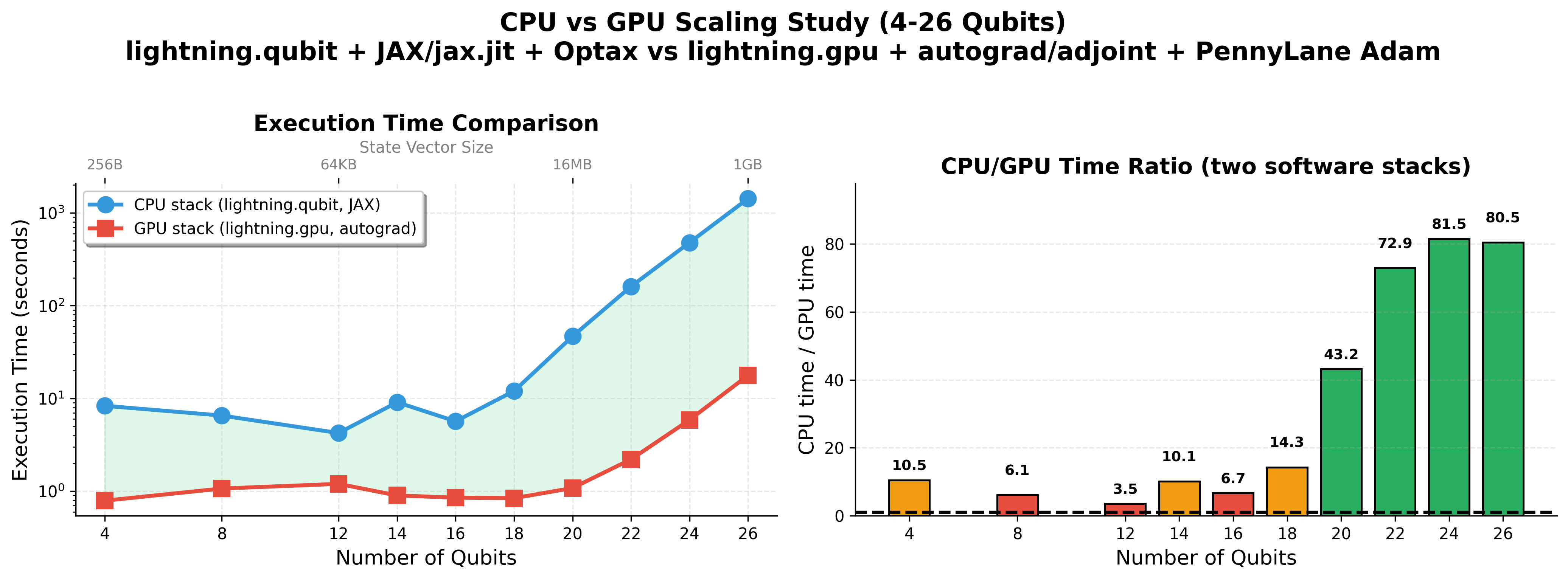}
    \caption{CPU and GPU stacks in the scaling study from 4 to 26 qubits. Left: runtime (log scale). Right: CPU/GPU time ratio.}
    \label{fig:scaling_study}
\end{figure}

To compare CPU and GPU simulation at larger sizes, we ran a separate VQE loop at 4 to 26 qubits (Table~\ref{tab:gpu_scaling} and Figure~\ref{fig:scaling_study}). Both paths use the same transverse-field Ising Hamiltonian with coupling $J = 1.0$ and field $h = 0.5$, and a hardware-efficient ansatz with two layers of $R_Y$ and $R_Z$ rotations followed by a CNOT chain. The GPU path has the lower runtime at every tested size, and the CPU/GPU time ratio ranges from 3.5 at 12 qubits to 81.5 at 24 qubits.

The two paths differ in more than the device. The CPU path uses \texttt{lightning.qubit} through the JAX interface with \texttt{jax.jit} and the Optax Adam optimizer, with JAX in its default 32-bit mode. The GPU path uses \texttt{lightning.gpu} through the autograd interface with adjoint differentiation and the PennyLane Adam optimizer. The ratio therefore compares two software stacks, not two devices. At 4 qubits the state vector holds 16 amplitudes, and we interpret the ratio of 10.5 at that size as overhead in the CPU stack rather than as a hardware effect.

\begin{table}[t]
    \centering
    \small
    \begin{tabular}{rrrrr}
        \toprule
        Qubits & State vector & CPU (s) & GPU (s) & CPU/GPU ratio \\
        \midrule
        4  & 256 B  & 8.33    & 0.79  & 10.5 \\
        8  & 4 KB   & 6.54    & 1.07  & 6.1  \\
        12 & 64 KB  & 4.24    & 1.20  & 3.5  \\
        14 & 256 KB & 9.10    & 0.90  & 10.1 \\
        16 & 1 MB   & 5.67    & 0.85  & 6.7  \\
        18 & 4 MB   & 12.03   & 0.84  & 14.3 \\
        20 & 16 MB  & 46.77   & 1.08  & 43.2 \\
        22 & 64 MB  & 161.07  & 2.21  & 72.9 \\
        24 & 256 MB & 478.73  & 5.87  & 81.5 \\
        26 & 1 GB   & 1425.06 & 17.71 & 80.5 \\
        \bottomrule
    \end{tabular}
    \caption{CPU and GPU runtimes in the scaling study. The state vector size is $2^n \times 16$ bytes (complex128). The ratio compares two software stacks that differ in more than the device (see text).}
    \label{tab:gpu_scaling}
\end{table}

\subsection{Multi-GPU Runs}

The multi-GPU benchmark uses a separate test problem: a sum of Pauli-$Z$ operators on $n$ qubits, a layer of $R_Y$ rotations (with an additional $R_Z$ layer in the maximum-size test) followed by a CNOT chain, \texttt{lightning.gpu} with adjoint differentiation, and gradient descent with step size 0.1. One MPI rank drives each GPU.

\begin{figure}[t]
    \centering
    \includegraphics[width=\textwidth]{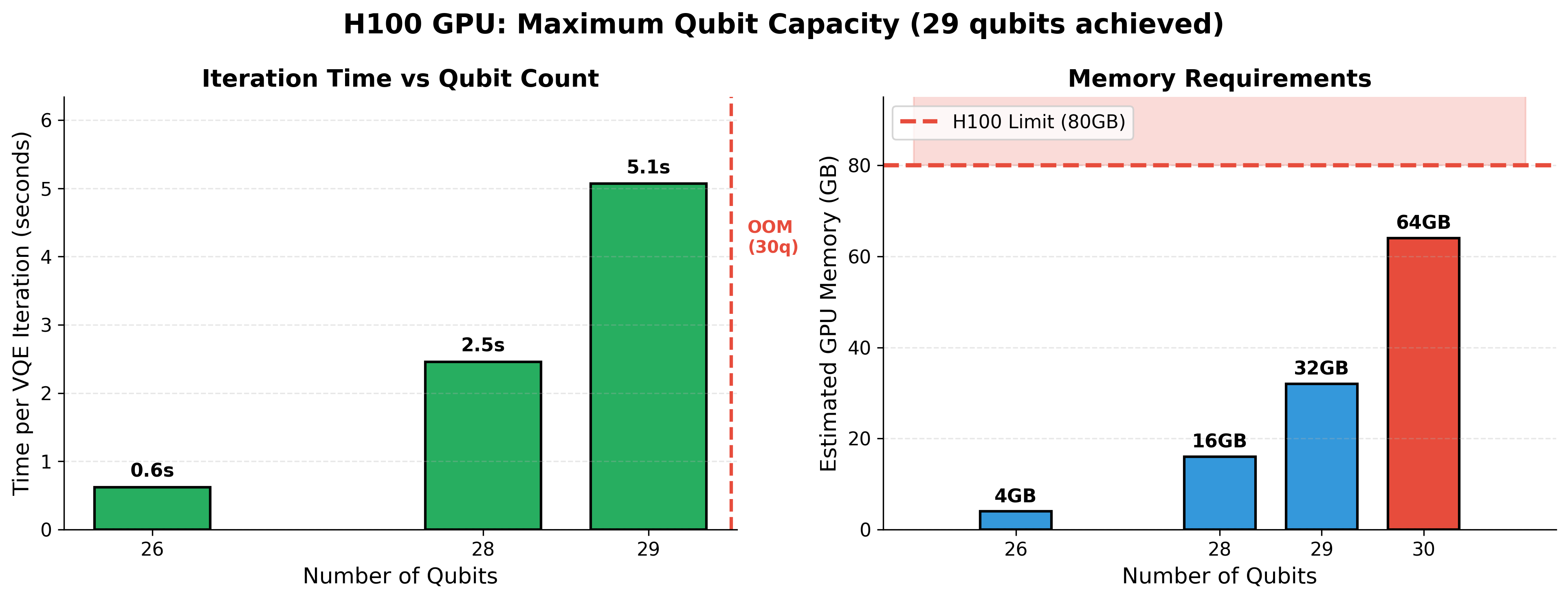}
    \caption{Maximum problem size on one NVIDIA H100. Left: measured time per iteration at 26, 28, and 29 qubits. Right: estimated GPU memory, four times the complex128 state vector. The 30-qubit run failed with an out-of-memory error.}
    \label{fig:multi_gpu}
\end{figure}

On one H100, 5 iterations at 26, 28, and 29 qubits take 0.62~s, 2.46~s, and 5.07~s per iteration, and the 30-qubit run fails with an out-of-memory error (Figure~\ref{fig:multi_gpu}). We estimate the GPU memory as four times the $2^n \times 16$-byte state vector. The estimate gives 64~GB at 30 qubits, below the 80~GB of the H100, so it undercounts the memory that \texttt{lightning.gpu} allocates.

In the load-balance test, each of the 4 GPUs solves 2 independent 22-qubit problems with 15 iterations each. The longest per-GPU time is 8.04~s, and the sum of the four per-GPU times is 31.99~s. Their ratio, 3.98, shows that the four GPUs carried equal load. It is not a measured speedup, because we did not run the 8 problems on a single GPU. In the throughput test, each GPU solves 3 independent 20-qubit problems, and the longest per-GPU time is 12.27~s.

\section{Discussion}
\label{sec:discussion}

The first question has a clear answer. In the STO-3G basis, the single-parameter double excitation ansatz can represent the exact ground state of H$_2$, and VQE reaches the exact energy at the equilibrium grid point to within $4\times10^{-6}$~Ha.

The answer to the second question is narrower than the one we set out to give. On one core, the largest runtime reduction came from doing fewer iterations, not from faster iterations: the stopping rule, the learning rate, and the warm start changed the runtime by a factor of 4.13 while the time per iteration stayed between 0.0198~s and 0.0204~s. At 4 qubits, \texttt{lightning.gpu} lowered the time per iteration to 0.0145~s but ran more iterations from cold starts and finished after the JIT version. The MPI runs completed the scan in 5.04~s to 8.45~s, but the serial and MPI jobs ran under different allocations, so our measurements do not isolate the parallel speedup. For 20 to 26 qubits, the GPU stack in the scaling study ran 43.2 to 81.5 times faster than the CPU stack, a comparison of two software stacks rather than of two devices. One H100 ran up to 29 qubits, and independent problems spread evenly across four GPUs, but without a single-GPU run of the same problems we cannot state a multi-GPU speedup.

These limits point to three practices for benchmarks of this kind. Each runtime should be reported with the optimizer, stopping rule, and starting point that produced it, because these settings changed the runtime by a factor of 4.13 in our runs. The serial baseline and the parallel runs should use one resource allocation, and the timer should cover the slowest rank. A multi-device efficiency needs a measured single-device run of the same work.

\section{Conclusions}

We implemented VQE for the H$_2$ potential energy surface and measured four implementations on the ERAU Vega cluster with 4 NVIDIA H100 GPUs. The single-parameter ansatz reproduces the exact STO-3G ground-state energy at equilibrium to within $4\times10^{-6}$~Ha. Early stopping, a larger learning rate, and warm starts reduce the iterations from 30,000 to 7,033 and the runtime from 593.95~s to 143.80~s. In the 4 to 26 qubit study, the CPU/GPU time ratio ranges from 3.5 to 81.5 for two software stacks that differ in interface, differentiation method, and optimizer. MPI runs with 2 to 32 ranks complete the scan in 8.45~s to 5.04~s, and one H100 runs up to 29 qubits. The bond-length scan is a parameter sweep in which each point is an independent optimization, and the same distribution across ranks or GPUs applies to other sweeps of this form, for example scans over several geometric parameters.

\section*{Changes in Version 2}

Version~1 of this preprint attributed a 4.13$\times$ runtime reduction to JIT compilation, reported an MPI speedup of 28.5$\times$ and a combined speedup of 117$\times$, and reported 99.4\% multi-GPU parallel efficiency. Version~2 corrects these statements. The 4.13$\times$ reduction tracks a 4.3$\times$ reduction in iteration count from early stopping, a larger learning rate, and warm starts. The MPI speedups compared jobs with different resource allocations, and version~2 withdraws them; the measured MPI runtimes remain. The 99.4\% figure divided the sum of per-GPU times by four times the longest per-GPU time, so it measured load balance, not speedup. Version~2 also corrects the description of the scaling-study Hamiltonian (transverse-field Ising, not a sum of Pauli-$Z$ operators), states the software differences between the CPU and GPU paths, and removes the figure built on the withdrawn speedups. It also corrects three entries of the Amdahl prediction table, the bond-length range (0.35 to 3.0~\AA, not 0.1 to 3.0~\AA), and the baseline iteration count (300, not 200), and it removes a preliminary serial table that used a smaller configuration than the cluster runs.

\section*{Acknowledgements}

We thank Dr. Khanal for guidance on parallelization strategies and HPC methodologies. This work used the ERAU Vega HPC cluster. The simulations used PennyLane with the Lightning backends, JAX for automatic differentiation, and Catalyst for JIT compilation. Large language model tools were used to assist with manuscript preparation and editing. The authors take full responsibility for all scientific content, methodology, and results.

\section*{Code and Data Availability}

The source code, job scripts, logs, and raw results are available at \url{https://code.rylanmalarchick.com/rylanmalarchick/QuantumVQE}, with an archived copy at \url{https://github.com/rylanmalarchick/QuantumVQE}.

\end{document}